# Graph Comparison Based on Adjacency Function Matrix

Arefe Alikhani[1], Farzad Didehvar[2]

[1, 2] Faculty of Mathematics and Computer Science, Amirkabir University of Technology, Tehran, Iran

a_alikhani@aut.ac.ir

didehvar@aut.ac.ir

# Abstract

In this paper, we present a new metric distance for comparing two large graphs to find similarities and differences between them based on one of the most important graph structural properties, which is 'Node Adjacency Information', for all vertices in the graph. Then, we defined a new function and some parameters to find the distance of two large graphs using different neighbors of vertices. There are some methods which they focused on the other features of graphs to obtain the distance between them, but some of them are 'Node Correspondence' which means their node set have the same size. However, in this paper, we introduce a new method which can find the distance between two large graphs with different size of node set.

## 1.1. Introduction

In the recent years, with the growth of data science, due to innovations in technology, data analyzing has become a frequent task in the fields such as social networks [1], economics, biology [2] and security systems [3], [4] and etc. In the last two decades, the research on networks has been increased dramatically due to predicting real-world phenomena using complex network models.

Thanks to the great growth of network models, comparing two large graphs is becoming more prevalent. Therefore, one of the frequent problems will be 'How we can find similarities and differences between two large graphs' [5], [6]. In other words, the main problem is 'Comparing Networks' for finding similarities and differences between them. It can be possible by grouping networks, appearing several clusters with various structures. As a result, the other problem which is so fundamental is 'How we can compare two different networks with respect to their structural properties such as Node Adjacency Information'. [7] For Comparing Networks, one of the most important things is 'Having an appropriate distance measure' [5], [6], which can differ two networks with huge differences and putting them in the same cluster if their graph based structures are so similar to each other.

Because of increase in the volume of data, the research on these topics in the recent years, was extended.

So, we face with a basic question for finding the best 'Distance Measure' for differentiating between two networks with various kinds of features.

There are many methods which have been defined as a 'Metric distance' and also some of them are not metric and they are just a measure for acquiring optimal answer as a distance between two networks [8], [6]. In this work, we present a new measure for finding the distance between large graphs using a function which contains the information of 'Node Adjacency' in the graph. The details of this new method will be discussed in the third chapter. During next chapter, some of the most used and popular methods for comparing two large graphs will be studied.

Throughout this work, we consider random graphs with different probabilities for their edges. Using distance measure, we can compare these random graphs with different structures as a model of real world networks [9]. Random graphs will be described in the third chapter.

## 2. Related Works:

For comparing two different networks, lots of distance measures were proposed until now. For each of them, some of particular features are important to consider. To compare two such structures, researchers often restrict their attention to two different categories of algorithms. [10]

In the first case, two networks have the same node sets which is called 'Known Node Correspondence (KNC)'. In other words, the matching between them is known. Some methods in this group are: Delta_con [11], [12] and Cut_Distance [13].

### 2.1. Cut Distance Method:

There is a relation between similarities of two complex networks or two large graphs and their cut distances [14]. Based on this fact, using cut distance, two directed or undirected graphs can be compared with high accuracy, according to their structural properties.

Let G=(V,E) be a weighted graph and C is a cut in this graph, then C, is able to partition the node set of the graph into two disjoint subsets S and T ( $S, T \subset V$ and C = (S,T) ) [14]

Therefore, for each Cut, there is a related 'cut-set', which contains edges having one endpoint in the sets S or T.

Cut-Set = $\{(u,v) \in E \mid u \in S, v \in T\}$

'Cut-Weight', is the other concept in this method, which is so important. Suppose that $w_{i,j}, i, j \in V$ is the edge weight in the graph G. Thus the total weight which crossed the Cut (C), is called 'Cut-Weight':

$$e_G(S,T) = \sum_{i \in S, j \in T} w_{ij}$$

The 'Cut-Distance' between two graphs G1=(V1,E1) and G2= (V2, E2) with the same vertex set is:

$$d(G1,G2) = \max \frac{1}{|V|} |e_{G1}(S,S^C) - e_{G2}(S,S^C)|$$

where, $s^C = $ V\S. [14]

Due to this definition, two large graphs are similar to each other, if their cut-distance is somehow close to each other and vice versa.

### 2.2. DeltaCon Method:

This method is able to find the similarity between two graphs on the same node set. Using similarity functions, found the optimize answer for this question: ' How can we find the similarity between two graphs?' [11]

The purpose of this problem is: 'Finding one Similarity Score between two graphs which is between 0 and 1'. Two graphs are totally different when the score is equal to 0 and they are exactly the same if the score value is 1 [15] , [16].

Although this method is very effective, but one of the drawbacks of that, is about the complexity and computability time. The computational complexity of the DeltaCon algorithm is quadratic respect to the number of nodes.

Second category of measures for comparing two graphs, is 'Unknown Node Correspondence (UNC)' and some of the popular methods in this category, are as follows:

### 2.3. Spectral methods:

In the Graph theory, spectrum has vast application in various areas. Because it carries fundamental information about structural properties of graphs. Based on this fact, for comparing the structure of two large graphs, we can use spectrum of the graphs. So this method is able to use information like eigenvectors and eigenvalues of the adjacency matrices of two graphs. [17], [18]

### 2.4. Graphlet-based methods:

In the recent years, information and data in various aspects of life, has been grown dramatically and thus, for analyzing and comparing data, we need to model them, so one of the powerful tools, are graphlets or motifs. Graphlets or motifs are highly recurrent patterns in the graphs and networks, that they are able to differ two graphs from structural point of view. [19]

### 3. Introducing new Measure

In this chapter, we introduce our new metric for 'Comparing Two Graphs' with different sizes. First of all, we need the population of random graphs with different probabilities for their edges which will be discussed by details in the following part:

Random Graphs can be expressed as a probability distribution over Graphs. 'Complex Network' and 'Modeling' are two fundamental fields, which random graphs play an important role in those areas [20].

### 3.2. Important Models of Random Graphs

Among different models for random graphs, G(n,m) and G(n,p) are two fundamental models which are closely related to 'Erdos-Renyi' model. [20]

### 3.3. Random Graphs as a First Population

Due to significant role of Random Graphs for modeling complex networks, we can use Random Graphs as a simulated data for running our algorithm.

However, all Random Graphs are not generated with the same probability 'p' for their edges. This is exactly the step for manual clustering using our input data.

### 3.4 . Our New Measure:

Let G be the set of different random graphs, that we generated using Erdos-Renyi model. Then in the first experiment it contains 'n'

random graphs with various probability in the various clusters.

$$G = \{ G_1, G_2, \ldots, G_n \}$$

In the $i^{th}$ random graph, $G_i = (V_i, E_i)$, suppose that 'α' be one of the vertices in the node set of $G_i$:

$$\alpha \in V(G_i)$$

**Def 1**:

For any random graph, $G_i \in G$, $N_{G_i,k}(\alpha)$ is the set of all k-neighbors for node 'α' in the graph $G_i$, which is as follows:

$$N_{G_i,1}(\alpha) =$$

$\{\beta \,|\, \beta \text{ is a node of } G_i \text{ and } (\alpha, \beta) \text{ is an edge in } G_i \}$

$$N_{G_i,k+1}(\alpha) =$$

$\{\beta \,|\, \beta \text{ is a node of } G_i \text{ and there exists}$
$\beta' \in N_{G_i,k}(\alpha) \text{ such that } (\beta, \beta') \in G_i\}$

### 3.5. Drawing a Line Graph for Each Node:

After finding the number of the four first neighbors for each node in the random graphs and defining parameters for them, next step would be, 'Drawing a Line Graph' based on the information exits in these vectors, to show the changes of the number of neighbors in different neighborhood.

**Def 2: Graph Neighbor α-Function**

$Diag(G_i, \alpha)$ is the diagram which is associated to the graph $G_i$ for each node α in this graph. We call this function, **'Graph Neighbor α-Function'**, for $\alpha \in V(G)$.

$$Diag(G_i, \alpha) =$$

$$\{ (k, |N_{G_i,k}|) : \ 1 \leq k \leq |V(G_i)| \}$$

Thus, $Diag(G_i)$ is the average of different values for $Diag(G_i, \alpha)$ such that $i \in \{1, 2, \ldots, n\}$. We call this function **'Graph Neighbor Function'** and it's definition is as follows:

$$Diag(G_i) =$$

$Avg(Diag(G_i, \alpha)) \text{ s.t } \alpha \in G_i =$

$$= \frac{1}{n} (\sum_{\alpha \in G_i} Diag(G_i, \alpha))$$

Using following relation, we can find the summation of values for $Diag(G_i, \alpha)$ and $Diag(G_i, \beta)$:

$$Diag(G_i, \alpha) + Diag(G_i, \beta) =$$

$$\{ \left( k, |N_{G_i,k}(\alpha)| + |N_{G_i,k}(\beta)| \right) : 1 \leq k \leq |V(G_i)| \}$$

### 3.6. Our Experimental results:

In our first experiment, we generated 100 random graphs using 'Erdos-Renyi' model and during this step, we used different probabilities for the edges of random graphs in various clusters. Thus the random graphs which exist in the same cluster have equal probabilities for their edges and vice versa. We expect the graphs in the same category, behave in the similar way. That means the distances between them after running our algorithm, should be close to each other, respect to the random graphs in another categories.

For 4 clusters that we considered, the probabilities are:

$$0.08, 0.22, 0.36, 0.5.$$

After implementing our algorithm on these random graphs, the results for the 4 neighbors of the first node, are as follows:

$$|N_{G_{1,1}(\alpha=1)}| = 9, \quad , |N_{G_{1,2}(\alpha=1)}| = 54$$

$$|N_{G_{1,3}(\alpha=1)}| = 98 \quad , |N_{G_{1,4}(\alpha=1)}| = 99$$

( i = 1  and k = 1,2,3,4 and α=1 )

Using these values, we can draw a line graph specifically for first node in the $G_1$, which is shown in the Figure 1:

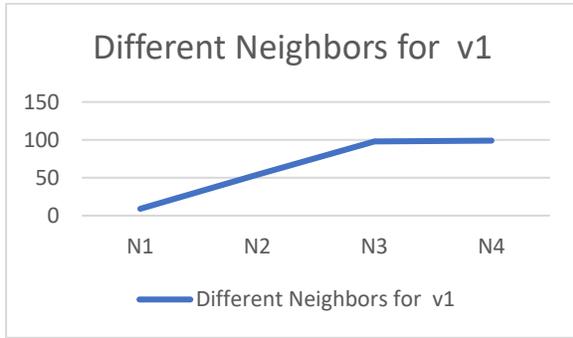

Figure 1 Different Neighbors for V1

Figure 2 displays Bar graphs for the exact number of different neighbors for first three nodes in the graph $G_1$.

Table 1 shows the exact number of different neighbors for nodes 1,2,3 in the random graph $G_1$.

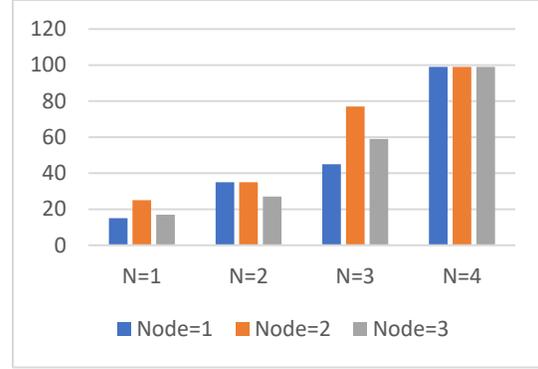

Figure 2 Number of neighbors for first 3 node in the random graph.

Table 1: Exact number of neighbors for first 3 nodes in the random graph

|     | Node=1 | Node=2 | Node=3 |
|-----|--------|--------|--------|
| N=1 | 15     | 25     | 17     |
| N=2 | 35     | 35     | 27     |
| N=3 | 45     | 77     | 59     |
| N=4 | 99     | 99     | 99     |

### 3.6. Finding the Average value for all 100 Random Graphs:

Our experiment includes 100 Random graphs and each of these graphs contains 100 nodes. As a result, it is required to find the number of four first neighbors of each node for first random graph (G1), then this procedure continues for the other random graphs (G2, G3, …, G100). After completing this steps, we have all the initial information for comparing two graphs. The other fundamental step would be 'Finding the Average value of $Diag(G_i, \alpha)$ which is defined as $Diag(G_i)$ ' the random graph $G_i$ and doing these steps for the other random graphs in the set 'G'.

## 3.7 Evaluating the method:

Therefore, for each random graph, after finishing all the previous steps, we would have one average value :

$$Diag_{G_1}, Diag_{G_2}, \ldots, Diag_{G_{100}}$$

for all 100 random graphs.

As a result, these 100 values are output of our method for comparing large networks. In the next section, we will analysis our results.

## 3.8. Distance Graph Neighbor Function:

**Def 3:**

Using all parameters and definitions in previous sections, the definition of '**Distance Graph Neighbor Function**' which is associated to the random graphs $G_i$ and $G_j$ is:

$$Diag\ Distance(G_i, G_j) = |Diag_{G_i} - Diag_{G_j}|$$

## 3.9 Results of our experiment :

In this experiment, we generated 100 random graphs based on Erdos-Renyi model and the probability of edges in these graphs are clustered into 4 groups.

| Clusters | Random Graphs | Probability of edges |
|---|---|---|
| Cluster1 | G1-G25 | 0.08 |
| Cluster2 | G26-G50 | 0.22 |
| Cluster3 | G51-G75 | 0.36 |
| Cluster4 | G76-G100 | 0.5 |

After running our algorithm on these four groups of random graphs, we found one value as a measure for comparing graphs. Different values for the four cluster of random graphs are shown as follows:

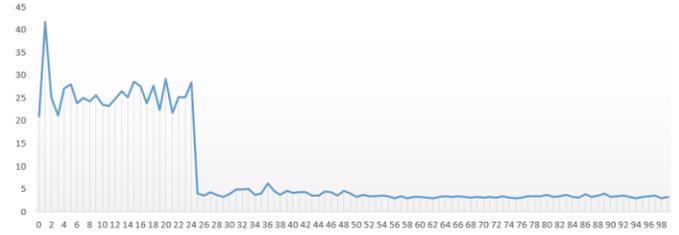

The outputs of algorithm from G25 to G26 is changing dramatically, compared to the other values for the other graphs.

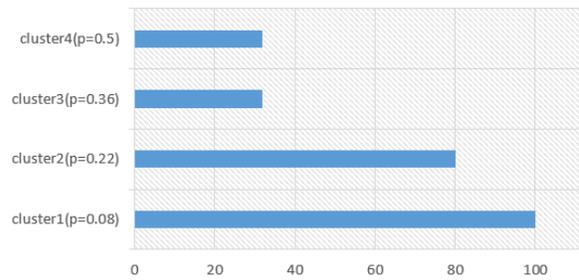

After evaluating output of our algorithm and comparing to the initial clusters that we generated them in the first step, we can find the accuracy of the algorithm. For the first cluster with p = 0.08, the accuracy of our algorithm for putting all graphs in correct category, is 100%. And for second cluster that the probability of graphs in this group is 0.22, the accuracy is 80%.

Also the time complexity of our algorithm is $O(|V_s| + |E_s|)$ that '$V_s$' and '$E_s$' are the number of vertices and edges in the $k^{th}$ neighborhood. Then it makes our method so efficient because it has proper time complexity respect to most of the methods which exist in the UNC category. Some of these methods have high accuracy in their results but their inappropriate complexity time, makes the implementation of their algorithm impossible.